\documentclass[letterpaper,journal]{IEEEtran}
\usepackage{amsmath,amsfonts}
\usepackage{algorithmic}
\usepackage{algorithm}
\usepackage{array}
\usepackage[caption=false,font=normalsize,labelfont=sf,textfont=sf]{subfig}
\usepackage{textcomp}
\usepackage{stfloats}
\usepackage{url}
\usepackage{verbatim}
\usepackage{graphicx}
\usepackage{cite}
\begin{document}

\title{Precision-Aware Hierarchical Spectrum Search for Finite-Precision MUSIC DOA Estimation}

\author{Linyun Yue, Chuang Yang,~\IEEEmembership{Member,~IEEE,} Jierui Chen, Mugen Peng,~\IEEEmembership{Fellow,~IEEE}
\thanks{This work was supported in part by the Beijing Natural Science Foundation under Grant L223007.
 (Corresponding author: Chuang Yang.) 
}
\thanks{The authors are with the Beijing Key Laboratory of Convergent Communications and Networking Technologies in LEO Satellite Systems, 
BeijingUniversity of Posts and Telecommunications, Beijing 100876, China (e-mail:
yuelinyun@bupt.edu.cn; chuangyang@bupt.edu.cn; badjerry0522@gmail.com;
derrickcr@bupt.edu.cn; pmg@bupt.edu.cn)}}



\maketitle

\begin{abstract}
Dense-grid spectrum search is computationally costly
for multiple signal classification (MUSIC)
direction-of-arrival (DOA) estimation, and low-precision
evaluation can introduce numerical errors that distort
candidate selection. This paper proposes precision-aware
hierarchical MUSIC (PAH-MUSIC), a multistage framework
that coordinates arithmetic precision, angular grids,
and candidate windows in spectrum evaluation, while
covariance estimation and eigendecomposition remain
in FP64. Stage-dependent precision is assigned from
coarse candidate discovery to final DOA estimation.
Under bounded perturbations and local spectral growth,
a conditional displacement bound gives a sufficient
window-width condition for retained, correctly associated
candidates. For a two-stage continuous cost model with
fixed precision weights and window coefficients, a
closed-form optimal coarse step and a cost-gain condition
are derived.
Simulations with two separated sources and 500 frames
at each tested signal-to-noise ratio (SNR) from
-10 to 15 dB show that FP8, FP16, and
FP32 coarse searches followed by FP64 refinement achieve
DOA root-mean-square errors close to a baseline using
FP64 at both search stages. Relative to that baseline,
the selected configurations reduce weighted
spectrum-evaluation cost by 28.36\%--44.66\% under the
adopted precision weights.
\end{abstract}


\section{Introduction}
\IEEEPARstart{D}{irection-of-arrival} (DOA) estimation serves as a key enabling 
technology for perception and localization across multiple domains. 
As sensing systems evolve from single-target detection toward the 
resolution of dense multi-target scenarios, super-resolution DOA 
estimation is increasingly important for automotive radar and
autonomous driving \cite{Bilik2019AutomotiveRadar}, as well as for
massive MIMO and future 6G sensing systems \cite{Liu2022ISAC}.
Subspace-based methods such as MUSIC \cite{Schmidt1986MUSIC}
and ESPRIT \cite{Roy1989ESPRIT} provide high-resolution DOA
estimation through subspace orthogonality and rotational
invariance, respectively.
However, MUSIC still requires eigendecomposition of the covariance 
matrix and spectral peak search over the entire angular domain, which 
two computational burdens render it a bottleneck for real-time processing.

Efforts to reduce the computational burden of MUSIC
have addressed three complementary aspects: the
computational formulation, subspace estimation, and
spatial spectrum search. At the formulation level,
real-valued methods replace complex-valued operations
with real-valued computations, reducing the arithmetic
cost of subspace decomposition and spectrum evaluation
\cite{Yan2014RealValued}. Beamspace processing projects
element-space observations onto a smaller set of beams,
reducing the processing dimension from the number of
array elements to the number of retained beams
\cite{Sun2004Beamspace}. At the subspace-estimation stage,
the propagator method (PM) and its variants exploit
linear relationships between partitions of the array
response to construct subspace information without
explicit eigendecomposition \cite{Marcos1995Propagator}.
Randomized methods instead
use matrix sketching and low-rank approximation to
estimate the dominant signal subspace at reduced cost
\cite{Li2021RandomizedMUSIC}. Randomized unitary MUSIC
(RU-MUSIC) combines this strategy with real-valued
processing \cite{Fang2026FinitePrecision}. At the
spectrum-search stage, approaches either eliminate
angular scanning or reduce the cost of spectrum
evaluation. Root-MUSIC replaces grid search with
polynomial rooting, although its polynomial degree
and associated rooting burden increase with the number
of array elements \cite{Barabell1983Root}. As a related
subspace-based alternative, ESPRIT avoids angular
scanning through rotational invariance but requires
an array structure supporting the necessary shift
invariance \cite{Roy1989ESPRIT}. Within grid-based MUSIC,
compressed search \cite{Yan2013Compressed}, fast Fourier
transforms \cite{Zhuang2017FFT}, and randomized sketching
with selective recomputation \cite{Li2021FastPseudospectrum}
reduce the search range or spectrum-evaluation cost.
Hierarchical methods further reduce the number of
evaluated angles by restricting fine-grid search to
coarse candidate regions, as in combined MUSIC for
coprime arrays \cite{Zhou2013DECOM}. For automotive
radar, Kim et al. proposed adaptive grid-refinement
MUSIC, using a coarse search over the full angular
range followed by fine searches within selected regions
of interest to reduce the search burden while retaining
high-resolution DOA estimation \cite{Kim2026AdaptiveGrid}.

Fang et al. developed blockwise mixed-precision and
adaptive precision strategies for RU-MUSIC, using
low-precision sums within blocks and higher-precision
accumulation across blocks \cite{Fang2026FinitePrecision}.
These strategies allocate arithmetic precision within
numerical computations. Hierarchical spectrum search
offers a further opportunity to allocate precision
across search stages, which serve different purposes.
The coarse stage identifies candidate regions for
subsequent refinement and therefore need not provide
accurate DOA estimates, provided that the relevant
spectral peaks remain within the retained regions.
This suggests using low-precision arithmetic for coarse
search to reduce computational cost, followed by
high-precision evaluation within the selected regions
for accurate DOA estimation. The key challenge is to
coordinate coarse-stage precision, angular grid spacing,
and candidate-window width so that computational savings
are achieved while retaining the peaks needed for
subsequent refinement.

Motivated by the different accuracy requirements of
coarse and fine searches, we propose precision-aware
hierarchical MUSIC (PAH-MUSIC), an efficient DOA
estimation framework that coordinates arithmetic
precision, angular grid spacing, and candidate-window
selection across search stages. Specifically,
low-precision arithmetic is used in the coarse stage
to identify candidate regions at reduced computational
cost, followed by higher-precision evaluation on finer
grids within these regions for accurate DOA estimation.
To guide candidate-window selection, we analyze the
effect of numerical perturbations on candidate locations
and derive a conditional displacement bound for
retained, correctly associated candidates. Furthermore,
we develop a two-stage spectrum-search cost model and
derive the optimal coarse-grid spacing and a condition
for cost reduction under prescribed precision weights
and window coefficients. Finally, simulations show that
coarse searches using FP8, FP16, or FP32 followed by
FP64 refinement achieve DOA root-mean-square errors
close to those of two-stage MUSIC using FP64 at both
stages, while reducing the weighted spectrum-evaluation
cost by 28.36\%--44.66\%.

\section{System Model and Finite-Precision Spectrum Evaluation}
\label{sec:system_model}

\subsection{Signal Model and MUSIC Spectrum Search}

Consider an $M$-element uniform linear array (ULA)
receiving signals from $N$ uncorrelated narrowband
far-field sources. The number of sources $N$ is assumed
known and satisfies $N<M$. The snapshot model is
\begin{equation}
    \mathbf{x}(t)
    = \mathbf{A}(\boldsymbol{\theta})\mathbf{s}(t)
    + \mathbf{n}(t),
    \label{eq:signal_model}
\end{equation}
where $\boldsymbol\theta=[\theta_1,\ldots,\theta_N]^T$
contains the DOAs,
$\mathbf A(\boldsymbol\theta)
=[\mathbf a(\theta_1),\ldots,\mathbf a(\theta_N)]$,
and $\mathbf s(t)$ contains the source signals.
The source-independent noise is
$\mathbf n(t)\sim\mathcal{CN}(\mathbf0,\sigma_n^2\mathbf I_M)$.
For spacing $d$, wavelength $\lambda$, and broadside
angle $\theta$, the unnormalized steering vector is
$[\mathbf{a}(\theta)]_r
=\exp[-\mathrm{j}2\pi(r-1)d\sin\theta/\lambda]$,
$r=1,\ldots,M$.

From $T$ snapshots, the FP64 sample covariance yields
$\widehat{\mathbf E}_{\mathrm n}\in\mathbb C^{M\times(M-N)}$,
containing the eigenvectors of its $M-N$ smallest
eigenvalues. The subscript $\mathrm n$ denotes noise.
The reference MUSIC denominator is
\begin{equation}
    D_{64}(\theta)
    =\|\widehat{\mathbf E}_{\mathrm n}^H
          \mathbf a(\theta)\|_2^2,
    \label{eq:music_denominator}
\end{equation}
whose minima correspond to peaks of
$P_{64}(\theta)=1/D_{64}(\theta)$.
All methods share FP64 covariance estimation and
eigendecomposition. Stage-dependent precision applies
only to the direct projection and squared-magnitude
summation in the spectrum search.

\subsection{Finite-Precision Spectrum Denominator Model}

Let $q=(p^{\mathrm{op}},p^{\mathrm{arith}})$ specify
operand and arithmetic formats. Quantization and
rounding in projection and squared-magnitude summation
\cite{Higham2022MixedPrecision} give, for finite values,
\begin{equation}
\begin{aligned}
\widehat D_q(\theta)
&=D_{64}(\theta)+e_q^{\mathrm{quant}}(\theta)
  +e_q^{\mathrm{arith}}(\theta)\\
&=D_{64}(\theta)+e_q(\theta),
\end{aligned}
\label{eq:fp_denominator}
\end{equation}
where $e_q^{\mathrm{quant}}$ measures operand quantization
with exact subsequent operations and $e_q^{\mathrm{arith}}$
is the additional arithmetic error. The common FP64
sample subspace shares sampling error across strategies;
$e_q$ describes numerical perturbations.

Operands and arithmetic use the same format: FP8
(E4M3FN), FP16, FP32, or FP64. Each real operation,
including sequential accumulation, is rounded in that
format, without fused multiply-add or block scaling.

Section~\ref{subsec:coverage_analysis} assumes finite
local perturbation bounds; nonfinite values lie outside
that model. Candidate displacement and omission during
selection are distinct failure mechanisms.

\subsection{Stage-Wise Precision Configuration and
Problem Formulation}
\label{subsec:stage_wise_precision}

Let $\Theta$ be the angular domain with length
$L_\theta$. An $n$-stage search, with $n\geq2$, uses grids
$\mathcal G_m$ with spacings
$s_1>\cdots>s_n=h>0$, where $h$ is the final
resolution. The stage formats are
\begin{equation}
    q_m=(b_m,b_m),\quad m=1,\ldots,n,
    \label{eq:precision_config}
\end{equation}
with $b_m\in\{\mathrm{FP8},\mathrm{FP16},
\mathrm{FP32},\mathrm{FP64}\}$ and
$b_1\preceq b_2\preceq\cdots\preceq b_n$,
where $\preceq$ denotes nondecreasing numerical
precision. The final format $b_n$ is configurable
according to the accuracy requirement.
Grid origins are fixed for each configuration.
Each stage obtains its operands from the original
FP64 subspace and steering vectors. Given the
retained region $\mathcal R_m$, it evaluates
$\mathcal G_m\cap\mathcal R_m$, with
$\mathcal R_1=\Theta$. A candidate at stage $m$
defines a next-stage window of total width
$W_{m+1}=\kappa_ms_m$ and half-width
$\kappa_ms_m/2$, where $\kappa_m>0$.
Overlapping windows are merged, and coincident
grid points are evaluated once per stage.

Let $G_m=|\mathcal G_m\cap\mathcal R_m|$ and
let $\rho_{b_m}>0$ be the per-point cost weight,
normalized by $\rho_{\mathrm{FP64}}=1$.
The precision-weighted spectrum-evaluation cost is
\begin{equation}
    C_{\mathrm{PAH}}
    =\sum_{m=1}^{n}\rho_{b_m}G_m.
    \label{eq:weighted_cost}
\end{equation}

The design objective is to reduce weighted search cost
while maintaining RMSE close to a specified FP64
two-stage search baseline:
\begin{equation}
\begin{aligned}
    \min_{\substack{\{b_m,s_m\}_{m=1}^{n}\\
                     \{\kappa_m\}_{m=1}^{n-1}}}
    \quad & C_{\mathrm{PAH}} \\
    \mathrm{s.t.}\quad
    & \mathrm{RMSE}_{\mathrm{PAH}}
      \leq \mathrm{RMSE}_{64,\mathrm{HS}}+\delta,
\end{aligned}
\label{eq:problem_formulation}
\end{equation}
where $\delta\geq0$ is an allowed RMSE degradation and
$\mathrm{RMSE}_{64,\mathrm{HS}}$ denotes the baseline
RMSE using FP64 at both stages, the same input data,
and the same final grid. The baseline configuration is
specified in Section~\ref{subsec:experimental_setup}.
Equation~\eqref{eq:problem_formulation} states a design
goal. Conditional local-minimum coverage
is analyzed separately in Section~\ref{subsec:coverage_analysis}.

\section{Proposed Precision-Aware Hierarchical MUSIC}
\label{sec:proposed_method}

\subsection{Precision-Aware Hierarchical Spectrum Search}
\label{subsec:hierarchical_search}

PAH-MUSIC specifies precision, grid spacing,
and candidate windows. For $m=1,\ldots,n$, the grid
is refined while precision is nondecreasing, and
stage $m$ evaluates
\begin{equation}
\begin{aligned}
    \Omega_m&=\mathcal G_m\cap\mathcal R_m,\\
    \widehat D_m(\theta)&=D_{64}(\theta)+e_m(\theta),
    \quad\theta\in\Omega_m,
\end{aligned}
\label{eq:stage_evaluation}
\end{equation}
where $\mathcal R_1=\Theta$. Operands are freshly
quantized from the original subspace and steering
vectors, with both operands and arithmetic in $b_m$.

Quantization may create multiple local minima within
one spectral basin. Ranking these minima alone can
assign several candidates to one source and omit
another; finer search cannot recover a reference
peak excluded from its search region. Thus, for
$m<n$, local peaks are detected within each connected
search interval, plateaus are represented by the grid
point nearest their midpoint, and representatives are
ranked by peak height. Greedy selection retains $N$
candidates subject to separation $\Delta_m\geq0$:
\begin{equation}
\begin{aligned}
    \mathcal C_m&=\{c_{m,1},\ldots,c_{m,N}\},\\
    |c_{m,i}-c_{m,j}|&\geq\Delta_m,\quad i\ne j.
\end{aligned}
\label{eq:candidate_collection}
\end{equation}
Setting $\Delta_m=0$ disables suppression; positive
values reduce nearby duplicate candidates but may
exclude closely spaced sources. Endpoints use
one-sided comparisons and ties favor smaller angles.
Finite, nonnegative denominators are eligible, with
zero assigned infinite peak height. Fewer than $N$
accepted candidates constitute a search failure.

Candidate $c_{m,k}$ defines the next-stage window
\begin{equation}
\begin{aligned}
    W_{m+1}&=\kappa_ms_m,\\
    I_{m+1,k}&=\left[c_{m,k}-\frac{W_{m+1}}2,
                         c_{m,k}+\frac{W_{m+1}}2\right]
                         \cap\Theta,
\end{aligned}
\label{eq:interval_expansion}
\end{equation}
where $W_{m+1}$ is the total width. The next search
region is
\begin{equation}
    \mathcal R_{m+1}=\operatorname{Merge}
    \left(\bigcup_{k=1}^{N}I_{m+1,k}\right).
\label{eq:region_propagation}
\end{equation}
Evaluation uses merged windows, while original windows
are retained for output association. At stage $n$,
$b_n$ and a fixed rule $\mathsf S_{\mathrm{out}}$
produce estimates, e.g., one minimum per window.
Empty required windows or insufficient valid outputs
constitute failure. Overlapping windows may return
repeated estimates, so coverage alone does not ensure
correct association.

\begin{algorithm}[!t]
\caption{Precision-Aware Hierarchical MUSIC}
\label{alg:pah_music}
\renewcommand{\algorithmicrequire}{\textbf{Input:}}
\renewcommand{\algorithmicensure}{\textbf{Output:}}
\begin{algorithmic}[1]
\REQUIRE $\widehat{\mathbf E}_{\mathrm n}$, $N$, $\Theta$;
         $\{\mathcal G_m,b_m\}_{m=1}^{n}$;
         $\{\kappa_m,\Delta_m\}_{m=1}^{n-1}$;
         $\mathsf S_{\mathrm{out}}$.
\ENSURE DOA estimates $\widehat{\boldsymbol\theta}$ or failure.
\STATE $\mathcal R_1\gets\Theta$.
\FOR{$m=1,\ldots,n$}
    \STATE $\Omega_m\gets\mathcal G_m\cap\mathcal R_m$.
    \IF{$\Omega_m=\emptyset$}
        \STATE \textbf{return} failure.
    \ENDIF
    \STATE $\widehat D_m\gets\operatorname{Evaluate}_{b_m}
           (\widehat{\mathbf E}_{\mathrm n},\Omega_m)$.
    \IF{$m<n$}
        \STATE $\mathcal C_m\gets\operatorname{SelectCandidates}
               (\widehat D_m,\Omega_m,N,\Delta_m)$.
        \IF{$|\mathcal C_m|<N$}
            \STATE \textbf{return} failure.
        \ENDIF
        \STATE Construct $\{I_{m+1,k}\}_{k=1}^{N}$
               using \eqref{eq:interval_expansion}.
        \STATE $\mathcal R_{m+1}\gets
               \operatorname{Merge}(\bigcup_{k=1}^{N}I_{m+1,k})$.
    \ELSE
        \STATE \textbf{if} a required window is empty,
               \textbf{return} failure.
        \STATE $\widehat{\boldsymbol\theta}\gets
               \mathsf S_{\mathrm{out}}
               (\widehat D_n,\Omega_n,\{I_{n,k}\}_{k=1}^{N},N)$.
        \STATE \textbf{if} fewer than $N$ valid outputs,
               \textbf{return} failure.
    \ENDIF
\ENDFOR
\STATE \textbf{return}
       $\operatorname{Sort}(\widehat{\boldsymbol\theta})$.
\end{algorithmic}
\end{algorithm}

\subsection{Conditional Candidate Displacement and Window Design}
\label{subsec:coverage_analysis}

Let $\theta_k^\star$ be a reference local minimum
of $D_{64}$ and $\theta^g_{k,m}$ its grid minimum
in a neighborhood $\mathcal U_{k,m}$ accessible at
stage $m<n$. For all
$\theta\in\mathcal U_{k,m}\cap\Omega_m$, assume
\begin{equation}
\begin{aligned}
 D_{64}(\theta)-D_{64}(\theta^g_{k,m})
 &\geq\mu_{k,m}(\theta-\theta^g_{k,m})^2,\\
 |e_m(\theta)|&\leq\epsilon_{k,m},
\end{aligned}
\label{eq:discrete_growth}
\end{equation}
where $\mu_{k,m}>0$ and $\epsilon_{k,m}\geq0$.
Let $g_{k,m}(s_m)$ bound
$|\theta^g_{k,m}-\theta_k^\star|$.

\noindent\textit{Proposition 1
(Conditional Candidate Displacement):}
If a retained candidate $c_{m,k}$ and
$\theta^g_{k,m}$ lie in
$\mathcal U_{k,m}\cap\Omega_m$ and satisfy
$\widehat D_m(c_{m,k})\leq
\widehat D_m(\theta^g_{k,m})$, then
\begin{equation}
 |c_{m,k}-\theta_k^\star|
 \leq g_{k,m}(s_m)
       +\sqrt{\frac{2\epsilon_{k,m}}{\mu_{k,m}}}.
\label{eq:candidate_displacement}
\end{equation}

\noindent\textit{Proof:}
The assumed comparison and error bound give
$D_{64}(c_{m,k})\leq D_{64}(\theta^g_{k,m})
+2\epsilon_{k,m}$. Combining this with
\eqref{eq:discrete_growth} bounds
$|c_{m,k}-\theta^g_{k,m}|$ by
$\sqrt{2\epsilon_{k,m}/\mu_{k,m}}$.
The triangle inequality gives
\eqref{eq:candidate_displacement}.
\hfill$\square$

Consequently, a sufficient next-window condition is
\begin{equation}
 \frac{\kappa_ms_m}{2}\geq g_{k,m}(s_m)
 +\sqrt{\frac{2\epsilon_{k,m}}{\mu_{k,m}}}
 +r_{k,m},
\label{eq:coverage_condition}
\end{equation}
where $r_{k,m}\geq0$ is a safety margin. Larger errors
or flatter spectra require wider windows. Taking
$g_{k,m}(s_m)\leq s_m/2$ requires an additional
grid-approximation assumption. Using
the bound at multiple stages requires its conditions
at every stage.

\subsection{Two-Stage Step-Size and Cost Analysis}
\label{subsec:complexity_analysis}

For the noise-subspace projection in
\eqref{eq:music_denominator}, the search complexity
is $\mathcal O(M(M-N)\sum_{m=1}^{n}G_m)$.
With $N$ windows per intermediate stage, ignoring overlap,
angular-boundary clipping, and discrete endpoint
counts gives the continuous approximation
\begin{equation}
 J^{(n)}=\rho_{b_1}\frac{L_\theta}{s_1}
 +N\sum_{m=2}^{n}\rho_{b_m}\kappa_{m-1}
                         \frac{s_{m-1}}{s_m}.
 \label{eq:multistage_cost}
\end{equation}
Actual costs use the unique-point counts in
\eqref{eq:weighted_cost}.

For the two-stage special case, let $s=s_1$,
$s_2=h$, and $\kappa=\kappa_1$. Denote the coarse
and final cost weights by $\rho_c$ and $\rho_f$.
Then
\begin{equation}
 J(s;\kappa)=\rho_c\frac{L_\theta}{s}
             +\rho_f\frac{N\kappa s}{h}.
 \label{eq:two_stage_cost}
\end{equation}
The first term decreases with coarse step size,
while the second increases through window expansion.

\noindent\textit{Proposition 2 (Fixed-Window-Coefficient Optimum):}
For fixed positive weights and $\kappa$, the
unconstrained continuous minimum is
\begin{equation}
\begin{aligned}
 s^\star&=\sqrt{\frac{\rho_cL_\theta h}
                         {\rho_fN\kappa}},\\
 J^\star&=2\sqrt{\frac{\rho_c\rho_fL_\theta N\kappa}{h}}.
\end{aligned}
 \label{eq:optimal_two_stage}
\end{equation}
\noindent\textit{Proof:}
For $A=\rho_cL_\theta$ and
$B=\rho_fN\kappa/h$, $A/s+Bs\geq2\sqrt{AB}$,
with equality at $A/s=Bs$.
\hfill$\square$

Compare with a two-stage baseline using the same
final format at both stages and window coefficient
$\kappa_{\mathrm{ref}}$. If both continuous optima
are feasible, their cost ratio and gain condition are
\begin{equation}
\begin{aligned}
 \frac{J^\star}{J_{\mathrm{ref}}^\star}
   &=\sqrt{\frac{\rho_c\kappa}
                      {\rho_f\kappa_{\mathrm{ref}}}},\\
 J^\star<J_{\mathrm{ref}}^\star
   &\iff\rho_c\kappa<\rho_f\kappa_{\mathrm{ref}}.
\end{aligned}
 \label{eq:two_stage_gain}
\end{equation}
The gain condition compares prescribed configurations
through both precision weight and window coefficient.
For the adopted FP64 final stage and
$\kappa_{\mathrm{ref}}=2$, the ratio is
$\sqrt{\rho_c\kappa/2}$.

For prescribed precision weights and window coefficients,
\eqref{eq:optimal_two_stage} provides a closed-form rule
for choosing the coarse-grid spacing. At $s^\star$, the
coarse and fine stages contribute equally to the weighted
search cost, balancing the number of coarse evaluations
against the extent of local refinement. Together with
\eqref{eq:two_stage_gain}, this result guides the selection
of efficient two-stage search configurations.

\section{Simulation Results and Discussion}
\label{sec:simulation}

\subsection{Experimental Setup}
\label{subsec:experimental_setup}

The experiments evaluate a two-stage instance of
PAH-MUSIC to assess finite-precision search in a
separated-source setting. A ULA with $M=20$ elements
and half-wavelength spacing ($d/\lambda=1/2$)
receives $N=2$ equal-power sources at
$-10.45^\circ$ and $10.88^\circ$, using $T=40$
snapshots. The sources are independent circularly
symmetric complex Gaussian signals,
$\mathbf s(t)\sim\mathcal{CN}(\mathbf0,\mathbf I_N)$,
with unit expected power
$\mathbb E[|s_k(t)|^2]=1$ for each source.
With the unit-magnitude steering-vector entries in
Section~\ref{sec:system_model}, the per-source SNR
at each array element is defined as
\begin{equation}
\begin{aligned}
 \mathrm{SNR}_{\mathrm{dB}}
 &=10\log_{10}\!\left(
       \frac{\mathbb E[|s_k(t)|^2]}{\sigma_n^2}\right)\\
 &=-10\log_{10}\sigma_n^2.
\end{aligned}
\label{eq:per_source_snr}
\end{equation}
The source-independent spatially white complex Gaussian
noise therefore has per-element variance
$\sigma_n^2=10^{-\mathrm{SNR}_{\mathrm{dB}}/10}$.
The search domain is
$[-90^\circ,90^\circ]$, and the final grid is
$-90^\circ:0.1^\circ:90^\circ$. Each per-source
SNR in $\{-10,-5,0,5,10,15\}$ dB uses 500
Monte Carlo frames, with identical input data and
the same FP64 covariance/EVD shared across strategies.

FP8 (E4M3FN), FP16, and FP32 coarse searches are
compared with an FP64 two-stage baseline; all use
FP64 final evaluation in this experiment. Candidate
selection uses $\Delta_1=5^\circ$, and each original
window returns one denominator minimum on the common
final grid. Table~\ref{tab:search_config} lists the
coarse-stage formats, window coefficients, coarse-grid
steps, and refinement-window widths used in the comparison.

\begin{table}[!t]
\caption{Two-Stage Search Configurations}
\label{tab:search_config}
\centering
\renewcommand{\arraystretch}{1.12}
\begin{tabular}{ccccc}
\hline
Coarse format & $\rho_{b_1}$ & $\kappa_1$
& $s_1$ ($^\circ$) & $W_2$ ($^\circ$) \\
\hline
FP8  & 0.125 & 5 & 0.474342 & 2.371708 \\
FP16 & 0.25  & 3 & 0.866025 & 2.598076 \\
FP32 & 0.5   & 2 & 1.500000 & 3.000000 \\
FP64 & 1     & 2 & 2.121320 & 4.242641 \\
\hline
\end{tabular}
\end{table}

The nominal per-evaluation weights are
\begin{equation}
 \begin{aligned}
 \rho_{\mathrm{FP8}}&=0.125, &\rho_{\mathrm{FP16}}&=0.25,\\
 \rho_{\mathrm{FP32}}&=0.5, &\rho_{\mathrm{FP64}}&=1.
 \end{aligned}
 \label{eq:nominal_weights}
\end{equation}
These weights are proportional to format bit widths
and normalized to FP64. They define a search-cost proxy
and do not measure hardware time or energy.

The window coefficients are selected using the
cost-gain condition in \eqref{eq:two_stage_gain}.
With FP64 final evaluation ($\rho_f=1$) and
$\kappa_{\mathrm{ref}}=2$, the break-even coefficient is
\begin{equation}
 \kappa_{\mathrm{eq}}(b_1)
 =\frac{\rho_f\kappa_{\mathrm{ref}}}{\rho_{b_1}}
 =\frac{2}{\rho_{b_1}}.
 \label{eq:break_even_coefficient}
\end{equation}
This gives $\kappa_{\mathrm{eq}}=4$, 8, and 16 for
FP32, FP16, and FP8, respectively. Choosing
$\kappa_1<\kappa_{\mathrm{eq}}$ and the corresponding
optimal coarse step in \eqref{eq:optimal_two_stage}
yields a lower continuous-model cost than the FP64
two-stage baseline. We select $\kappa_1=2$, 3, and 5
for FP32, FP16, and FP8, respectively, within these
cost-saving ranges. All configurations are fixed across
SNRs. The subsequent RMSE comparison evaluates whether
these configurations retain estimation accuracy while
reducing search cost.

\begin{figure}[!t]
    \centering
    \includegraphics[width=\columnwidth]{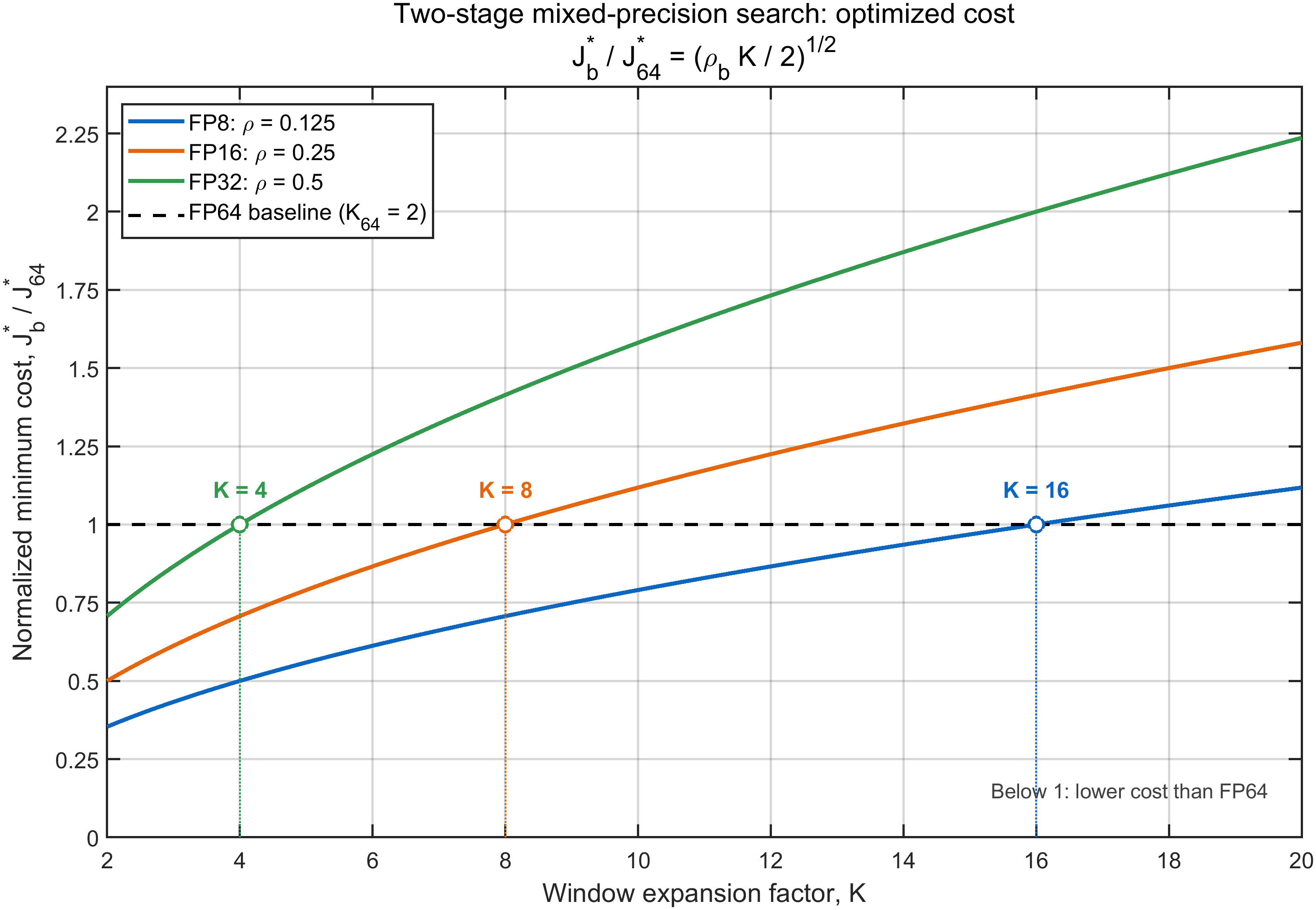}
    \caption{Continuous-model cost ratio versus window
    coefficient ($K$ in the plot denotes $\kappa$).
    The coarse step is optimized separately at each
    coefficient; final evaluation is FP64 and the
    FP64 two-stage baseline uses $\kappa_{\mathrm{ref}}=2$.
    The curves do not impose candidate-coverage constraints.}
    \label{fig:cost_boundary}
\end{figure}

\subsection{DOA Estimation and Candidate Coverage}
\label{subsec:accuracy_recovery}

\begin{figure}[!t]
    \centering
    \includegraphics[width=\columnwidth]{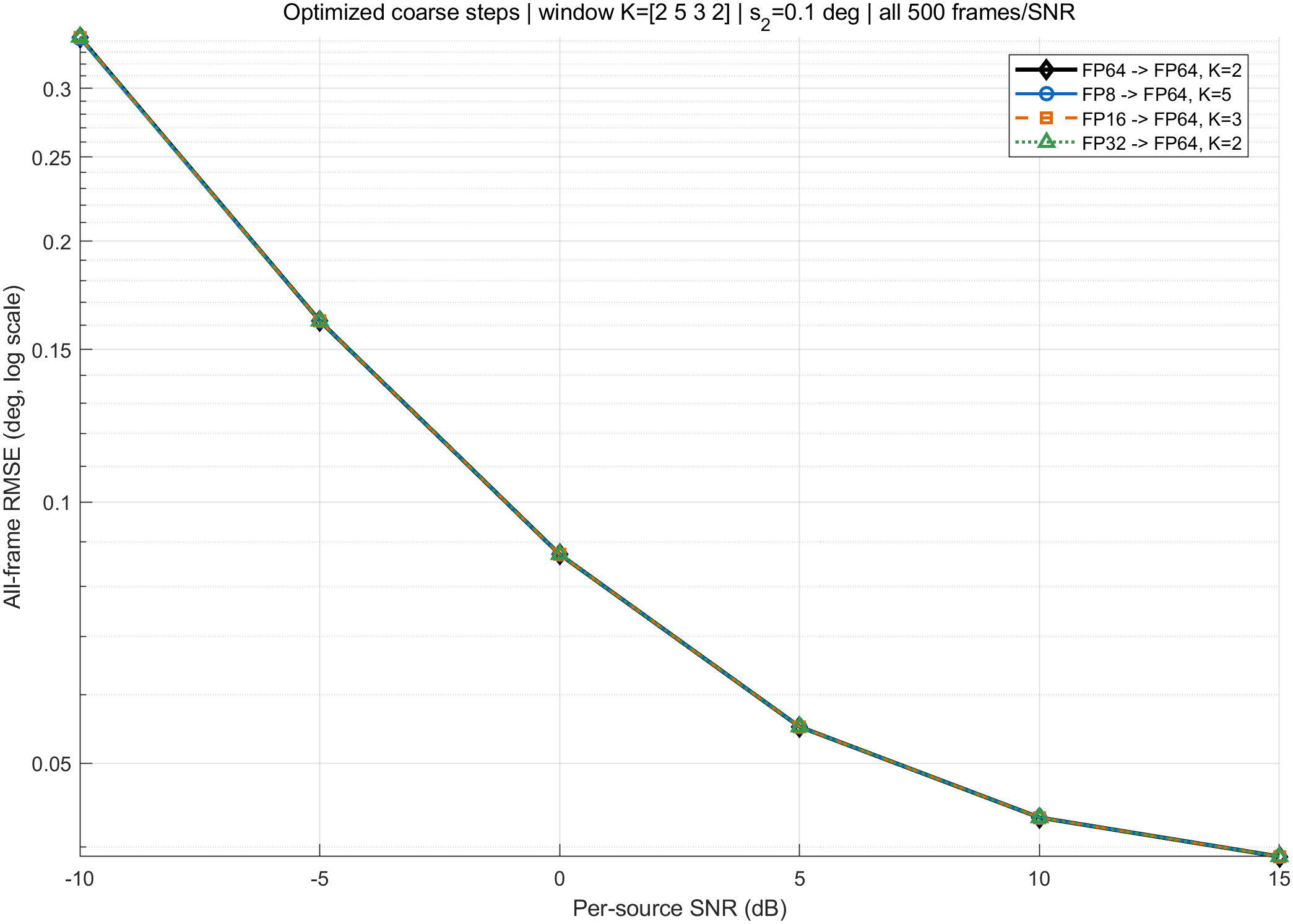}
    \caption{All-frame DOA RMSE versus per-source SNR
    for the configurations in Table~\ref{tab:search_config},
    using 500 frames per SNR. All strategies use FP64
    final evaluation. The baseline is two-stage FP64
    MUSIC; $K$ in the legend denotes $\kappa_1$.}
    \label{fig:accuracy_recovery}
\end{figure}

Estimated DOAs are matched one-to-one to the true
directions by minimizing the sum of squared angular
errors over all permutations. For this one-dimensional
criterion, pairing sorted angle lists attains the
minimum. With $N_{\mathrm{MC}}=500$ and
$e_{i,k}$ denoting the resulting error for source $k$
in frame $i$, the all-frame RMSE is
\begin{equation}
 \mathrm{RMSE}=\sqrt{\frac{1}{N_{\mathrm{MC}}N}
             \sum_{i=1}^{N_{\mathrm{MC}}}
             \sum_{k=1}^{N}e_{i,k}^2}.
 \label{eq:simulation_rmse}
\end{equation}
All four strategies return complete outputs in the
reported frames. Fig.~\ref{fig:accuracy_recovery}
shows closely overlapping RMSE curves. At $-10$ dB,
FP8 yields $0.342707^\circ$, compared with
$0.343753^\circ$ for the FP64 two-stage baseline,
showing similar error levels in this sampled scenario.
The close agreement in RMSE verifies the feasibility
of the selected configurations: coefficients below the
theoretical cost thresholds allow low-precision candidate
discovery followed by accurate FP64 refinement in the
tested setting.

To distinguish similar aggregate RMSE from identical
outputs, estimates are also compared frame by frame
with the FP64 two-stage outputs. The paired output
difference RMSE is zero for FP16 and FP32 at all
six SNRs. For FP8, it is zero from $-5$ to 15 dB
and $0.015492^\circ$ at $-10$ dB.

True-source joint window coverage is the fraction
of frames in which all true DOAs lie in
$\mathcal R_2$. At $-10$ dB, it is 94.8\%, 99.2\%,
and 98.4\% for FP8, FP16, and FP32, respectively.
An estimated peak may remain inside a window even
when the true DOA lies just outside, so this diagnostic
can differ from paired output agreement without a large
RMSE change. This true-DOA diagnostic differs from
the conditional local-minimum coverage in Proposition~1.

\subsection{Spectrum-Search Cost and Model Comparison}
\label{subsec:cost_analysis}

Fig.~\ref{fig:cost_boundary} illustrates
\eqref{eq:two_stage_gain} for the adopted weights.
The FP8, FP16, and FP32 curves intersect the FP64
baseline at $\kappa=16$, 8, and 4, respectively.
The selected coefficients 5, 3, and 2 lie below their
respective break-even thresholds, giving predicted cost
ratios of 0.5590, 0.6124, and 0.7071. Together with
the RMSE results in Fig.~\ref{fig:accuracy_recovery},
these ratios show how the theoretical cost criterion
guides the construction of efficient search configurations.

\begin{figure}[!t]
    \centering
    \includegraphics[width=\columnwidth,
        trim=0 0 557.4bp 22.5bp,clip]{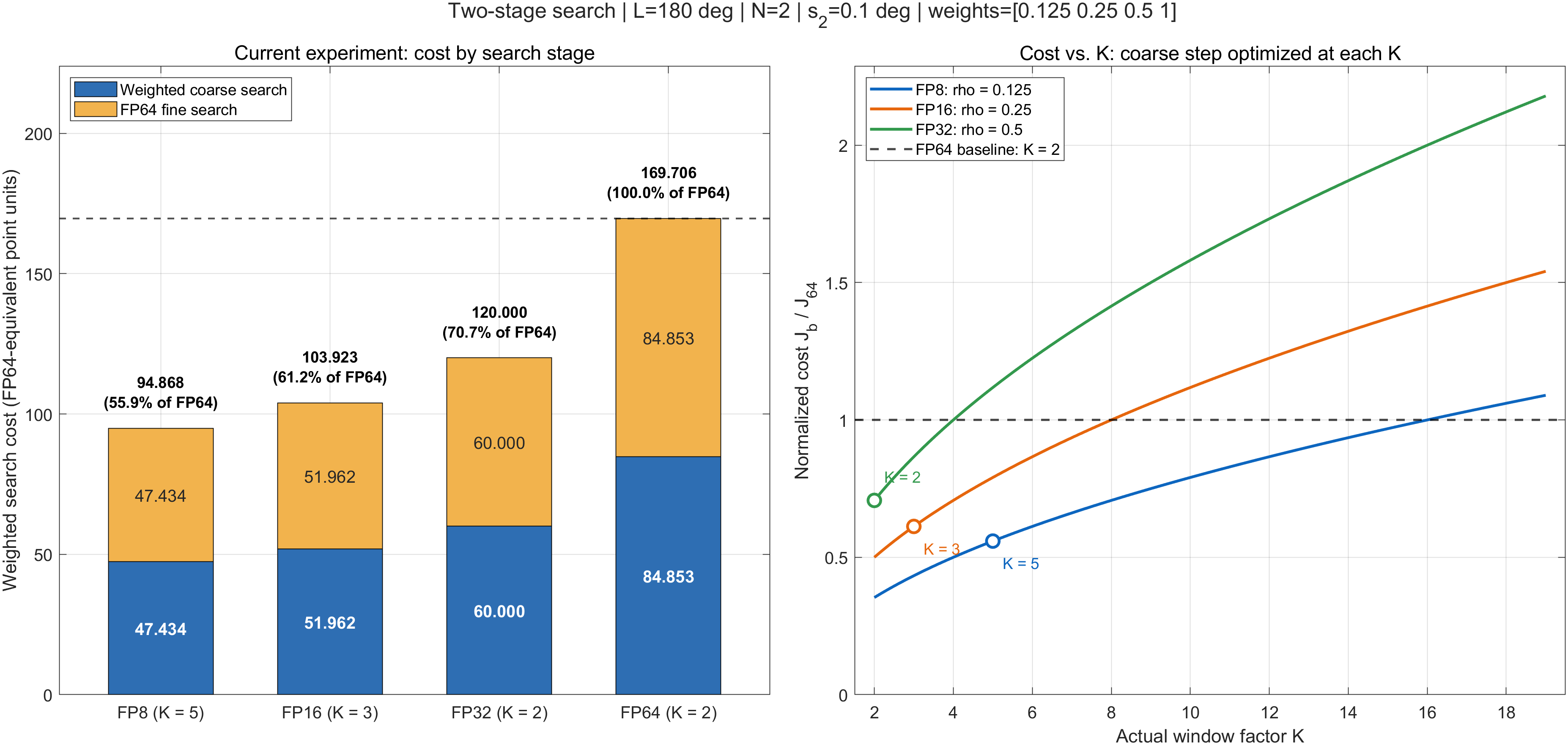}
    \caption{Continuous-model coarse/fine cost decomposition
    for Table~\ref{tab:search_config}. Both stage contributions
    are equal at the analytic optimum. Bar heights are
    model predictions, expressed in FP64-equivalent
    evaluation units, rather than counted evaluations.}
    \label{fig:cost_comparison}
\end{figure}

Fig.~\ref{fig:cost_comparison} gives theoretical
costs of 94.8683, 103.9230, and 120.0000 for FP8,
FP16, and FP32, compared with 169.7056 for the
FP64 baseline. The corresponding reductions are
44.10\%, 38.76\%, and 29.29\%. Equal coarse/fine
contributions illustrate the balance underlying
Proposition~2.

For actual grid evaluation, let $G_{c,b}$ and
$G_{f,b}$ count the distinct coarse and fine points
for coarse format $b$. The weighted count is
\begin{equation}
 C_b=\rho_bG_{c,b}+G_{f,b},
 \label{eq:pah_experimental_cost}
\end{equation}
where fine points in overlapping windows are counted
once. Relative to the same-SNR mean FP64 two-stage
weighted count, the observed reductions across the
six SNRs are 44.54\%--44.66\% for FP8,
39.04\%--39.16\% for FP16, and
28.36\%--28.51\% for FP32. Small differences from
the continuous model reflect discrete grid and
endpoint counting.

These selected configurations provide additional
weighted-cost savings over FP64 two-stage search at
similar observed RMSE in the tested separated-source
setting.

\section{Conclusion}
\label{sec:conclusion}

PAH-MUSIC coordinates stage-wise precision, angular
grids, and candidate windows in MUSIC spectrum search.
A conditional displacement bound relates local numerical
perturbations to window coverage. For two-stage search,
the continuous model yields an optimal coarse step and
a cost-gain condition at fixed precision weights and
window coefficients. In the tested separated two-source
scenario over $-10$ to 15 dB, the selected FP8, FP16,
and FP32 coarse-search configurations with FP64 refinement
achieved RMSE close to FP64 two-stage search while
reducing weighted spectrum-evaluation cost by
28.36\%--44.66\%.

\end{document}